\documentclass[journal]{IEEEtran}%
\usepackage{amsmath}
\usepackage{amsfonts}
\usepackage{amssymb}
\usepackage{graphicx}
\usepackage{amsthm}
\usepackage{cite}
\usepackage{verbatim}
\usepackage{multicol}%
\providecommand{\U}[1]{\protect\rule{.1in}{.1in}}
\newtheorem{theorem}{Theorem}

\newtheorem{lemma}{Lemma}

\newtheorem{remark}{Remark}
\begin{document}

\title{Robust Distributed Control Protocols for Large Vehicular Platoons with
Prescribed Transient and Steady State Performance}
\author{Christos~K.~Verginis,~Charalampos~P.~Bechlioulis,~Dimos~V.~Dimarogonas~and~Kostas~J.~Kyriakopoulos\thanks{C.
K. Verginis and D. V. Dimarogonas are with the Centre for Autonomous Systems
at Kungliga Tekniska Hogskolan, Stockholm 10044, Sweden. C. P. Bechlioulis and
K. J. Kyriakopoulos are with the Control Systems Laboratory, School of
Mechanical Engineering, National Technical University of Athens, Athens 15780,
Greece. Emails:{{ cverginis@kth.se, chmpechl@mail.ntua.gr, dimos@kth.se,
kkyria@mail.ntua.gr}}.}\thanks{This work was supported by the EU funded
project \ RECONFIG: Cognitive, Decentralized Coordination of Heterogeneous
Multi-Robot Systems via Reconfigurable Task Planning (FP7-ICT-600825,
2013-2016), the Swedish Research Council (VR) and the Knut and Alice
Wallenberg foundation.}}
\maketitle

\begin{abstract}
In this paper, we study the longitudinal control problem for a platoon of
vehicles with unknown nonlinear dynamics under both the predecessor-following
and the bidirectional control architectures. The proposed control protocols
are fully distributed in the sense that each vehicle utilizes feedback from
its relative position with respect to its preceding and following vehicles as
well as its own velocity, which can all be easily obtained by onboard sensors.
Moreover, no previous knowledge of model nonlinearities/disturbances is
incorporated in the control design, enhancing in that way the robustness of
the overall closed loop system against model imperfections. Additionally,
certain designer-specified performance functions determine the transient and
steady-state response, thus preventing connectivity breaks
due to sensor limitations as well as inter-vehicular collisions. Finally, extensive simulation studies and a
real-time experiment conducted with mobile robots clarify the proposed control
protocols and verify their effectiveness.

\end{abstract}


\IEEEpeerreviewmaketitle

\section{Introduction\label{Sec:Introduction}}

\IEEEPARstart{D}{uring} the last few decades, Automated Highway Systems (AHS)
have drawn a notable amount of attention in the field of automatic control.
Unlike human drivers, that are not able to react quickly and accurately enough
to follow each other in close proximity at high speeds, the safety and
capacity of highways (measured in vehicles/lanes/time) is significantly
increased when vehicles operate autonomously forming large platoons at close spacing.

Guaranteed string stability \cite{1996Swaroop_Hedrick} was first achieved via
centralized control schemes
\cite{2005Jovanovic_Bamieh,2008Qu_Wang_Hull,2001Soo_Chong_Roh}, with all
vehicles either communicating explicitly with each other or sending
information to a central computer that determined the control protocol. To
enhance the overall system's autonomy and avoid delay problems due to wireless
communication \cite{2001Liu_Goldsmith_Mahal_Hedrick}, decentralized schemes
were developed, dealing either with the predecessor-following (PF)
architecture
\cite{2000Stankovic_Stanojevic_Siljak,2000Rajamani_et_al,1999Liang_Ying_Peng},
where each vehicle has access to its relative position with respect to its
preceding vehicle, or the bidirectional (BD) architecture
\cite{2013Hao_Barooah,2009Barooah_Mehta_Hespanha,2012Lin_Fardad_Jovanovic},
where each vehicle measures its relative position with respect to its
following vehicle as well. Furthermore, in a few works
\cite{2001Liu_Goldsmith_Mahal_Hedrick ,2012Guo_Yue} a combined predecessor and
leader-following architecture was developed according to which each vehicle
obtains additional information from the leading vehicle. Finally,
\cite{Zheng2015,Zheng2016} addressed various architectures by examining
different kinds of information flow topologies.

The majority of the works in the related literature either considers linear
vehicle dynamic models and controllers
\cite{1999Liang_Ying_Peng,2006Yadlapalli_et_al,2004Seiler_Pant_Hedrick} or
adopt linearization techniques and Linear Quadratic optimal control
\cite{2000Stankovic_Stanojevic_Siljak,Zheng2015,Zheng2016,2013Hao_Barooah}.
However, linearization may lead to unstable inner dynamics since the estimated
linear models deviate in general from the real ones, away from the
corresponding linearization points. In particular, a comparison of the
aforementioned control architectures was carried out in
\cite{2004Seiler_Pant_Hedrick}, where it was stated that double integrator
models with linear controllers under the predecessor-following architecture
may lead to string instability. String instability conditions were also
presented in \cite{Middleton2010}. Finally, in
\cite{1994Swaroop_Hedrick_Chien_Ioannou} a comparison of two common control
policies was conducted; namely the constant time headway policy and the
constant spacing policy, that are related to the inter-vehicular distances of
the platoon. Particularly for the latter, it was also stated that feedback
from the leading vehicle needs to be constantly broadcasted.

Another important issue associated with the decentralized control of large
platoons of vehicles, concerns the fact that in many works the transient and
steady state response of the closed loop system is affected severely by the
control gains' selection and the number of vehicles as stated in
\cite{2009Barooah_Mehta_Hespanha,2012Lin_Fardad_Jovanovic,2004Seiler_Pant_Hedrick}%
, limiting thus the controller's capabilities. Furthermore, the majority of
the results on the aforementioned decentralized architectures consider known
(either partially of fully) dynamic models and parameters, which may lead to
poor closed loop performance in the presence of parametric uncertainties and
unknown external disturbances.

In this work, we propose decentralized control protocols for large platoons of
vehicles with $2^{nd}$ order\footnote{The results may be easily extended for
$3^{rd}$ order dynamics, that model the driving/braking force with a
first-order inertial transfer function \cite{Zheng2015,Zheng2016}, following
similar design steps as presented in \cite{2014Bechlioulis_Rovithakis}.
However, in this work, we adopted a $2^{nd}$ order model only to present more
clearly the control design philosophy and highlight its properties.} uncertain
nonlinear dynamics, under both the predecessor-following and the bidirectional
control architectures. The desired feasible formation is created arbitrarily
fast and is maintained with arbitrary accuracy avoiding simultaneously any
connectivity breaks (owing to limited sensor capabilities) and inter-vehicular collisions. 
The developed schemes exhibit the following significant
characteristics. First, they are purely distributed in the sense that the
control signal of each vehicle is calculated based solely: a) on local
relative position information with respect to its preceding and following
vehicles, as well as b) on its own velocity, both of which can be easily
acquired by its on-board sensors. Furthermore, their complexity proves to be
considerably low. Very few and simple calculations are required to output the
control signals. Additionally, they do not require any previous knowledge of the
vehicle's dynamic model parameters and no estimation models are employed to
acquire such knowledge. Moreover, contrary to the related works, the transient
and steady state response is fully decoupled by: i) the number of vehicles
composing the platoon, ii) the control gains selection and iii) the vehicle
model uncertainties. In particular, the achieved performance as well as the
collision avoidance and the connectivity maintenance are a priori and
explicitly imposed by certain designer-specified performance functions, thus
simplifying significantly the selection of the control gains. Tuning of the
controller gains is only confined to achieving reasonable control effort.

\section{Problem Statement \label{Sec:Problem_Statement}}

We consider the longitudinal formation control problem of $N$ vehicles with
$2^{nd}$ order nonlinear dynamics:
\begin{align}
\dot{p}_{i} &  =v_{i}\nonumber\\
m_{i}\dot{v}_{i} &  =f_{i}(v_{i})+u_{i}+w_{i}(t)\text{, }i=1,\ldots
,N\label{eqn:nonlinear_dynamics}%
\end{align}
where $p_{i}$ and $v_{i}$ denote the position and velocity of each vehicle
respectively, $m_{i}$ is the mass, which is considered unknown, $f_{i}(v_{i})$
is an unknown continuous nonlinear function that models the aerodynamic
friction (drag), $u_{i}$ is the control input and $w_{i}(t)$ is a bounded
piecewise continuous function of time representing exogenous disturbances. The
control objective is to design a distributed control protocol such that a
rigid formation is established with prescribed transient and steady state
performance, despite the presence of model uncertainties. By prescribed
performance, we mean that the formation is achieved in a predefined transient
period and is maintained arbitrarily accurate while avoiding connectivity breaks 
and collisions with neighboring vehicles. The geometry of the formation
is represented by the desired gaps $\Delta_{i-1,i}$, $i=1,\ldots,N$ between
consecutive vehicles, where $\Delta_{i-1,i}>0$ denotes the desired distance
between the $(i-1)$-th and $i$-th vehicle. In general, all $\Delta_{i-1,i}$
are given as control specifications and are directly related to the platoon
velocity as well as to the input constraints of the traction/bracking forces
(for safety reasons). Moreover, the inter-vehicular distance $p_{i-1}%
(t)-p_{i}(t)$, $i=1,\dots,N$ should be kept greater than $\Delta
_{\mathsf{col}}$ to avoid collisions and less than $\Delta_{\mathsf{con}}$ to
maintain the network connectivity owing to the limited sensing capabilities of
the vehicles (e.g., when employing range sensors to measure the distance
between two successive vehicles). Furthermore, to ensure the feasibility of
the desired formation, we assume that $\Delta_{\mathsf{col}}<\Delta
_{i-1,i}<\Delta_{\mathsf{con}}$, $i=1,\ldots,N$. Additionally, the reference
command of the formation is generated by a leading vehicle with position
$p_{0}(t)$ and bounded velocity $v_{0}(t)$. Finally, to solve the
aforementioned formation control problem, the following assumption is required.

\textbf{Assumption A1. }The initial state of the platoon does not violate the
collision and connectivity constraints. That is $\Delta_{\mathsf{col}}<$
$p_{i-1}(0)-p_{i}(0)<\Delta_{\mathsf{con}},$ $i=1,\ldots,N$.

In this work, we consider two distributed control architectures: (a) the
predecessor-following (PF) architecture, according to which the control action
of each vehicle is based only on its preceding vehicle and (b) the
bidirectional (BD) architecture, where the control action of each vehicle
depends on the information from both its preceding and following
vehicles. Hence, let us formulate the control variables $e_{p_{i}}%
(t)=p_{i-1}(t)-p_{i}(t)-\Delta_{i-1,i}$, $i=1,\ldots,N$. Equivalently, the
neighborhood error vector $e_{p}\triangleq\lbrack e_{p_{1}},\ldots,e_{p_{N}%
}]^{T}$ may be expressed as follows:
\begin{equation}
e_{p}=Se_{p_{0}}\label{eqn:error_vector}%
\end{equation}
where $e_{p_{0}}\triangleq\lbrack e_{p_{0,1}},\ldots,e_{p_{0,N}}]^{T}=\bar
{p}_{0}-p-\bar{\Delta}_{0}$ is the error with respect to the leading vehicle,
$p\triangleq\lbrack p_{1},\ldots,p_{N}]^{T}\in\Re^{N},$ $\bar{p}_{0}%
\triangleq\lbrack p_{0},\ldots,p_{0}]^{T}\in\Re^{N}$, $\bar{\Delta}%
_{0}\triangleq\lbrack\Delta_{0,1},\Delta_{0,2},\ldots,\Delta_{0,N}]^{T}\in
\Re^{N}$ with $\Delta_{0,i}=\sum_{j=1}^{i}\Delta_{j-1,j},$ $i=1,\ldots,N$ and
$S\in\Re^{N\times N}=\left[  s_{i,j}\right]  $ where $s_{i,i}=1$,
$s_{i+1,i}=-1$ and $s_{i,j}=0$ for all other elements, with $i,j=1,\dots,N$.
Notice that $S$ has strictly positive singular values \cite{2008Qu_Wang_Hull}
and since all principal minors of $S$ are equal to $1,$ $S$ is also a
nonsingular $\mathcal{M}$-matrix\footnote{An $\mathcal{M}$-matrix is a square
matrix whose off-diagonal elements are less than or equal to zero and whose
eigenvalues have positive real part.} \cite{1994Berman_Plemmons}. Finally, the
following lemma regarding nonsingular $\mathcal{M}$-matrices will be
employed to derive the main results of this paper.

\begin{lemma}
\label{Lemma;M-matrix}\cite{2008Qu_Wang_Hull} Consider a nonsingular
$\mathcal{M}$-matrix $A\in\Re^{N\times N}$. There exists a diagonal positive
definite matrix $P=$ $(\mathrm{diag}(A^{-1}\underline{\mathbf{1}}))^{-1}$ such
that $PA+A^{T}P$ is positive definite.
\end{lemma}

\section{Main Results\label{sec:Main-Results}}

In this work, prescribed performance will be adopted in order: i) to achieve
predefined transient and steady state response for each neighborhood position
error $e_{p_{i}}(t)$, $i=1,\ldots,N$ as well as ii) to avoid the violation of
the collision and connectivity constraints as presented in Section
\ref{Sec:Problem_Statement}.

\subsection{Sufficient Conditions\label{subsec:Sufficient-Condition}}

Prescribed performance is achieved when the neighborhood position errors
$e_{p_{i}}(t)$, $i=1,\ldots,N$ evolve strictly within predefined regions that
are bounded by absolutely decaying functions of time, called performance
functions \cite{2008Bechlioulis_Rovithakis,2014Bechlioulis_Rovithakis}. In
this work, the mathematical expression of prescribed performance is formulated
by the following inequalities:
\begin{equation}
-\underline{M}_{p_{i}}\rho_{p_{i}}\left(  t\right)  <e_{p_{i}}\left(
t\right)  <\overline{M}_{p_{i}}\rho_{p_{i}}\left(  t\right)  \text{, }\forall
t\geq0 \label{eqn:pp_inequalities}%
\end{equation}
for all $i=1,\ldots,N$, where:
\begin{equation}
\rho_{p_{i}}\left(  t\right)  =\left(  1-\tfrac{\rho_{\infty}}{\max
\{\underline{M}_{p_{i}},\overline{M}_{p_{i}}\}}\right)  \exp\left(
-lt\right)  +\tfrac{\rho_{\infty}}{\max\{\underline{M}_{p_{i}},\overline
{M}_{p_{i}}\}} \label{eqn:performance_functions}%
\end{equation}
are designer-specified, smooth, bounded and decreasing functions of time with
$l$, $\rho_{\infty}$ positive parameters incorporating the desired transient
and steady state performance specifications respectively, and $\underline{M}%
_{p_{i}},$ $\overline{M}_{p_{i}}$, $i=1,\ldots,N$ positive parameters selected
appropriately to satisfy the collision and connectivity constraints, as
presented in the sequel. In particular, the decreasing rate of $\rho_{p_{i}%
}\left(  t\right)  ,$ $i=1,\ldots,N$, which is affected by the constant $l$,
introduces a lower bound on the speed of convergence of $e_{p_{i}}(t),$
$i=1,\ldots,N$. Furthermore, depending on the accuracy of the measurement
device, the constant $\rho_{\infty}$ can be set arbitrarily small
$\rho_{\infty}\ll\underline{M}_{p_{i}},\overline{M}_{p_{i}}$, $i=1,\dots,N$,
thus achieving practical convergence of $e_{p_{i}}(t)$, $i=1,\ldots,N$ to
zero. Additionally, we select:
\begin{equation}
\underline{M}_{p_{i}}=\Delta_{i-1,i}-\Delta_{\mathsf{col}}\And\overline
{M}_{p_{i}}=\Delta_{\mathsf{con}}-\Delta_{i-1,i}\text{, }i=1,\dots,N.
\label{eqn:M_upper+lower}%
\end{equation}
Apparently, since the desired formation is compatible with the collision and
connectivity constraints (i.e., $\Delta_{\mathsf{col}}<\Delta_{i-1,i}%
<\Delta_{\mathsf{con}},$ $i=1,\dots,N$), \eqref{eqn:M_upper+lower} ensures
that $\underline{M}_{p_{i}},$ $\overline{M}_{p_{i}}>0,$ $i=1,\dots,N$ and
consequently under \textbf{Assumption A1} (i.e., $\Delta_{\mathsf{col}}<$
$p_{i-1}(0)-p_{i}(0)<\Delta_{\mathsf{con}},$ $i=1,\ldots,N$) that:
\begin{equation}
-\underline{M}_{p_{i}}\rho_{p_{i}}\left(  0\right)  <e_{p_{i}}\left(
0\right)  <\overline{M}_{p_{i}}\rho_{p_{i}}\left(  0\right)  \text{,
}i=1,\dots,N. \label{eqn:pp_t_zero}%
\end{equation}
Hence, guaranteeing prescribed performance via (\ref{eqn:pp_inequalities}) for
all $t>0$ and employing the decreasing property of $\rho_{p_{i}}\left(
t\right)  $, $i=1,\dots,N,$ we obtain $-\underline{M}_{p_{i}}<e_{p_{i}}\left(
t\right)  <\overline{M}_{p_{i}},$ $\forall t\geq0$ and consequently, owing to
(\ref{eqn:M_upper+lower}), $\Delta_{\mathsf{col}}<p_{i-1}(t)-p_{i}%
(t)<\Delta_{\mathsf{con}}$ for all $t\geq0$ and $i=1,\ldots,N$, which ensures
collision avoidance and connectivity maintenance for all $t\geq0$. Therefore,
imposing prescribed performance via (\ref{eqn:pp_inequalities}) with
appropriately selected performance functions $\rho_{p_{i}}\left(  t\right)  $,
$i=1,\dots,N$ and positive constant parameters $\underline{M}_{p_{i}%
},\overline{M}_{p_{i}}$, $i=1,\dots,N$, as dictated in
(\ref{eqn:performance_functions}) and (\ref{eqn:M_upper+lower}) respectively,
proves sufficient to solve the robust formation control problem stated in
Section \ref{Sec:Problem_Statement}.

\subsection{Control Design\label{subsec:Control-Design}}

\noindent\textbf{Kinematic Controller:} Given the neighborhood position errors
$e_{p_{i}}\left(  t\right)  =p_{i-1}(t)-p_{i}(t)-\Delta_{i-1,i}$,
$i=1,\ldots,N$:

\textit{Step I-a.} Select the corresponding functions $\rho_{p_{i}}\left(
t\right)  $ and positive parameters $\underline{M}_{p_{i}},$ $\overline
{M}_{p_{i}}$, $i=1,\ldots,N$ following (\ref{eqn:performance_functions}) and
(\ref{eqn:M_upper+lower}) respectively, in order to incorporate the desired
transient and steady state performance specifications as well as the collision
and connectivity constraints.

\textit{Step I-b.} Define the normalized position errors as:
\begin{equation}
\xi_{p}(e_{p},t)\triangleq%
\begin{bmatrix}
\xi_{p_{1}}(e_{p_{1}},t)\\
\vdots\\
\xi_{p_{N}}(e_{p_{N}},t)
\end{bmatrix}
\triangleq%
\begin{bmatrix}
\frac{e_{p_{1}}}{\rho_{p_{1}}\left(  t\right)  }\\
\vdots\\
\frac{e_{p_{N}}}{\rho_{p_{N}}\left(  t\right)  }%
\end{bmatrix}
\equiv(\rho_{p}\left(  t\right)  )^{-1}e_{p}\text{,} \label{eqn:ksi_p}%
\end{equation}
where $\rho_{p}\left(  t\right)  =\mathrm{diag}\left(  [\rho_{p_{i}}\left(
t\right)  ]_{i=1,\ldots,N}\right)  $ as well as the control signals:
\begin{align}
r_{p}(\xi_{p})  &  =\mathrm{diag}\left(  \left[  \tfrac{\frac{1}%
{\underline{M}_{p_{i}}}+\frac{1}{\overline{M}_{p_{i}}}}{(1+\frac{\xi_{p_{i}}%
}{\underline{M}_{p_{i}}})(1-\frac{\xi_{p_{i}}}{\overline{M}_{p_{i}}})}\right]
_{i=1,\dots,N}\right) \label{eqn:r_p}\\
\varepsilon_{p}(\xi_{p})  &  =\left[  \ln\left(  \tfrac{1+\frac{\xi_{p_{1}}%
}{\underline{M}_{p_{1}}}}{1-\frac{\xi_{p_{1}}}{\overline{M}_{p_{1}}}}\right)
,\ldots,\ln\left(  \tfrac{1+\frac{\xi_{p_{N}}}{\underline{M}_{p_{N}}}}%
{1-\frac{\xi_{p_{N}}}{\overline{M}_{p_{N}}}}\right)  \right]  ^{T}\text{.}
\label{eqn:epsilon_p}%
\end{align}

\textit{Step I-c.} Design the reference velocity vector for the
predecessor-following and bidirectional control architectures as follows:

A. Predecessor-Following architecture:
\begin{align}
v_{d}(\xi_{p},t)  &  \triangleq%
\begin{bmatrix}
v_{d_{1}}(\xi_{p_{1}},t)\\
\vdots\\
v_{d_{N-1}}(\xi_{p_{N-1}},t)\\
v_{d_{N}}(\xi_{p_{N}},t)
\end{bmatrix}
\nonumber\\
&  =k_{p}(\rho_{p}\left(  t\right)  )^{-1}r_{p}(\xi_{p})\varepsilon_{p}%
(\xi_{p})\text{,} \label{eqn:v_d_predecessor_f}%
\end{align}

B. Bidirectional architecture:
\begin{align}
v_{d}(\xi_{p},t)  &  \triangleq%
\begin{bmatrix}
v_{d_{1}}(\xi_{p_{1}},\xi_{p_{2}},t)\\
\vdots\\
v_{d_{N-1}}(\xi_{p_{N-1}},\xi_{p_{N}},t)\\
v_{d_{N}}(\xi_{p_{N}},t)
\end{bmatrix}
\nonumber\\
&  =k_{p}S^{T}(\rho_{p}\left(  t\right)  )^{-1}r_{p}(\xi_{p})\varepsilon
_{p}(\xi_{p}) \label{eqn:v_d_bidirectional}%
\end{align}
with $k_{p}>0$.

\noindent\textbf{Dynamic Controller:}

\textit{Step II-a.} Define the velocity error vector $e_{v}\triangleq\lbrack
e_{v_{1}},\ldots,e_{v_{N}}]^{T}=v-v_{d}(\xi_{p},t)$ with $v\triangleq\lbrack
v_{1},\ldots,v_{N}]^{T}$ for both control architectures and select the
corresponding velocity performance functions $\rho_{v_{i}}\left(  t\right)  ,$
$i=1,\ldots,N$ such that $\rho_{v_{i}}\left(  0\right)  >\left\vert e_{v_{i}%
}(0)\right\vert $, $i=1,\ldots,N$.

\textit{Step II-b.} Similarly to the first step define the normalized velocity
errors as:
\begin{equation}
\xi_{v}(e_{v},t)\triangleq%
\begin{bmatrix}
\xi_{v_{1}}(e_{v_{1}},t)\\
\vdots\\
\xi_{v_{N}}(e_{v_{N}},t)
\end{bmatrix}
\triangleq%
\begin{bmatrix}
\frac{e_{v_{1}}}{\rho_{v_{1}}\left(  t\right)  }\\
\vdots\\
\frac{e_{v_{N}}}{\rho_{v_{N}}\left(  t\right)  }%
\end{bmatrix}
\equiv(\rho_{v}\left(  t\right)  )^{-1}e_{v}\text{,} \label{eqn:ksi_v}%
\end{equation}
where $\rho_{v}\left(  t\right)  =\mathrm{diag}\left(  [\rho_{v_{i}}\left(
t\right)  ]_{i=1,\ldots,N}\right)  $ as well as the control signals:
\begin{align}
r_{v}(\xi_{v})  &  =\mathrm{diag}\left(  \left[  \tfrac{2}{(1+\xi_{v_{i}%
})(1-\xi_{v_{i}})}\right]  _{i=1,\dots,N}\right) \label{eqn:r_v}\\
\varepsilon_{v}(\xi_{v})  &  =\left[  \ln\left(  \tfrac{1+\xi_{v_{1}}}%
{1-\xi_{v_{1}}}\right)  ,\ldots,\ln\left(  \tfrac{1+\xi_{v_{N}}}{1-\xi_{v_{N}%
}}\right)  \right]  ^{T}\text{.} \label{eqn:epsilon_v}%
\end{align}

\textit{Step II-c.} Design the distributed control protocol for both
architectures as follows:
\begin{equation}
u(\xi_{v},t)\triangleq%
\begin{bmatrix}
u_{1}(\xi_{v_{1}},t)\\
\vdots\\
u_{N}(\xi_{v_{N}},t)
\end{bmatrix}
=-k_{v}(\rho_{v}\left(  t\right)  )^{-1}r_{v}(\xi_{v})\varepsilon_{v}(\xi_{v})
\label{eqn:u}%
\end{equation}
with $k_{v}>0$.

\begin{remark}
(Control Philosophy) The prescribed performance control technique guarantees the prescribed transient and steady state performance specifications, that are encapsulated in (\ref{eqn:pp_inequalities}), by enforcing the normalized position
errors $\xi_{p_{i}}\left(  t\right)  $ and velocity errors $\xi_{v_{i}}\left(
t\right)  $, $i=1,\dots,N$ to remain strictly within the sets $\left(
-\underline{M}_{p_{i}},\overline{M}_{p_{i}}\right)  $ and $\left(
-1,1\right)  $ respectively for all $t\geq0$. Notice that modulating
$\xi_{p_{i}}\left(  t\right)  $ and $\xi_{v_{i}}\left(  t\right)  $ via the
logarithmic functions $\ln\left(  \tfrac{1+\frac{\star}{\underline{M}_{p_{i}}%
}}{1-\frac{\star}{\overline{M}_{p_{i}}}}\right)  $ and $\ln\left(
\tfrac{1+\star}{1-\star}\right)  $ in the control signals (\ref{eqn:epsilon_p}%
), (\ref{eqn:epsilon_v}) and selecting $\underline{M}_{p_{i}}$, $\overline
{M}_{p_{i}}$ according to (\ref{eqn:M_upper+lower}) and $\rho_{v_{i}}\left(
0\right)  >\left\vert e_{v_{i}}(0)\right\vert $, the control signals
$\varepsilon_{p}(\xi_{p})$ and $\varepsilon_{v}(\xi_{v})$ are initially well
defined. Moreover, it is not difficult to verify that maintaining simply the
boundedness of the modulated errors $\varepsilon_{p}(\xi_{p}\left(  t\right)
)$ and $\varepsilon_{v}(\xi_{v}\left(  t\right)  )$ for all $t\geq0$ is
equivalent to guaranteeing $\xi_{p_{i}}\left(  t\right)  \in\left(
-\underline{M}_{p_{i}},\overline{M}_{p_{i}}\right)  $ and $\xi_{v_{i}}\left(
t\right)  \in\left(  -1,1\right)  $ for all $t\geq0$. Therefore, the problem
at hand can be visualized as stabilizing the modulated errors $\varepsilon
_{p}(\xi_{p}\left(  t\right)  )$ and $\varepsilon_{v}(\xi_{v}\left(  t\right)
)$, within the feasible regions defined via $\xi_{p_{i}}\in\left(
-\underline{M}_{p_{i}},\overline{M}_{p_{i}}\right)  $ and $\xi_{v_{i}}%
\in\left(  -1,1\right)  $, $i=1,\dots,N$ for all $t\geq0$. A careful
inspection of the proposed control scheme (\ref{eqn:v_d_predecessor_f}),
(\ref{eqn:v_d_bidirectional}) and (\ref{eqn:u}) reveals that it actually
operates similarly to barrier functions in constrained optimization, admitting
high negative or positive values depending on whether $e_{p_{i}}\left(
t\right)  \rightarrow-\underline{M}_{p_{i}}\rho_{p_{i}}\left(  t\right)  $ and
$e_{v_{i}}\left(  t\right)  \rightarrow\rho_{v_{i}}\left(  t\right)  $ or
$e_{p_{i}}\left(  t\right)  \rightarrow\overline{M}_{p_{i}}\rho_{p_{i}}\left(
t\right)  $ and $e_{v_{i}}\left(  t\right)  \rightarrow-\rho_{v_{i}}\left(
t\right)  $, $i=1,\dots,N$ respectively; eventually preventing $e_{p_{i}%
}\left(  t\right)  $ and $e_{v_{i}}\left(  t\right)  $, $i=1,\dots,N$ from
reaching the corresponding boundaries.
\end{remark}


\begin{remark}
(Selecting the Performance Functions) Regarding the construction of the performance functions, we stress that the
desired performance specifications concerning the transient and steady state
response as well as the collision and connectivity constraints are introduced
in the proposed control schemes via $\rho_{p_{i}}\left(  t\right)  $ and
$\underline{M}_{p_{i}}$, $\overline{M}_{p_{i}}$, $i=1,\dots,N$ respectively.
In addition, the velocity performance functions $\rho_{v_{i}}\left(  t\right)
$ impose prescribed performance on the velocity errors
$e_{v_{i}}=v_{i}-v_{d_{i}}$, $i=1,\dots,N$. In this respect, notice that
$v_{d_{i}}$ act as reference signals for the corresponding
velocities $v_{i}$, $i=1,\dots,N$. However, it should be stressed that
although such performance specifications are not required (only the
neighborhood position errors need to satisfy predefined transient and steady
state performance specifications) their selection affects both the evolution
of the position errors within the corresponding performance envelopes as well
as the control input characteristics (magnitude and rate). Nevertheless, the
only hard constraint attached to their definition is related to their initial
values. Specifically, $\rho_{v_{i}}\left(  0\right)$ should
be chosen to satisfy $\rho_{v_{i}}\left(  0\right)  >\left\vert e_{v_{i}%
}(0)\right\vert $, $i=1,\ldots,N$.
\end{remark}

\subsection{Stability Analysis\label{subsec:STABILITYANALYSIS}}

The main results of this work are summarized in the following theorem, where
it is proven that the aforementioned distributed control protocols solve the
robust formation problem with prescribed performance under collision and
connectivity constraints for the considered platoon of vehicles.

\begin{theorem}
\label{th:main_theorem} Consider a platoon of $N$ vehicles with uncertain 2nd
order nonlinear dynamics (\ref{eqn:nonlinear_dynamics}), that aims at
establishing a formation described by the desired inter-vehicular gaps
$\Delta_{i-1,i}$, $i=1,\ldots,N$, while satisfying the collision and
connectivity constraints represented by $\Delta_{\mathsf{col}}$ and
$\Delta_{\mathsf{con}}$ respectively with $\Delta_{\mathsf{col}}%
<\Delta_{i-1,i}<\Delta_{\mathsf{con}}$, $i=1,\ldots,N$. Under
\textbf{Assumption A1}, the distributed control protocols (\ref{eqn:ksi_p}%
)-(\ref{eqn:u}), for the predecessor-following and bidirectional control
architectures, guarantee: $-\underline{M}_{p_{i}}\rho_{p_{i}}\left(  t\right)  <e_{p_{i}}\left(t\right)  <\overline{M}_{p_{i}}\rho_{p_{i}}\left(  t\right)  \text{, }$ for all $t\geq0$ and $i=1,\ldots,N$, as well as the boundedness of all closed loop signals.
\end{theorem}

\emph{Proof:} The proof of Theorem \ref{th:main_theorem} proceeds in three
phases. First, we show that $\xi_{p_{i}}\left(  t\right)  $ and $\xi_{v_{i}%
}\left(  t\right)  $ remain within $\left(  -\underline{M}_{p_{i}}%
,\overline{M}_{p_{i}}\right)  $ and $\left(  -1,1\right)  $ respectively, for
a specific time interval $\left[  0,\tau_{\max}\right)  $ (i.e., existence of
a maximal solution). Next, we prove that the proposed control scheme retains
$\xi_{p_{i}}\left(  t\right)  $ and $\xi_{v_{i}}\left(  t\right)  $ strictly
in compact subsets of $\left(  -\underline{M}_{p_{i}},\overline{M}_{p_{i}%
}\right)  $ and $\left(  -1,1\right)  $ for all $t\in\left[  0,\tau_{\max
}\right)  $, which by contradiction leads to $\tau_{\max}=\infty$
(i.e., forward completeness) in the last phase, thus completing the proof. It
should be also noticed that the proof is provided only for the
predecessor-following architecture since the proof for the bidirectional case
follows identical steps.

In particular, by differentiating (\ref{eqn:ksi_p}) and (\ref{eqn:ksi_v}), we obtain:
\begin{align}
\dot{\xi}_{p} &  =(\rho_{p}\left(  t\right)  )^{-1}(\dot{e}_{p}-\dot{\rho}%
_{p}\left(  t\right)  \xi_{p})\label{eqn:ksi_p_dot}\\
\dot{\xi}_{v} &  =(\rho_{v}\left(  t\right)  )^{-1}(\dot{e}_{v}-\dot{\rho}%
_{v}\left(  t\right)  \xi_{v})\label{eqn:ksi_v_dot}%
\end{align}
Employing (\ref{eqn:nonlinear_dynamics}), (\ref{eqn:error_vector}) as well as
the fact that $v_{i}\equiv v_{d_{i}}+\rho_{v_{i}}\left(  t\right)  \xi_{v_{i}%
}$ and substituting (\ref{eqn:v_d_predecessor_f}), (\ref{eqn:u}) in
(\ref{eqn:ksi_p_dot}) and (\ref{eqn:ksi_v_dot}), we arrive at:
\begin{align}
\dot{\xi}_{p} &  =h_{p_{A}}(t,\xi)\nonumber\\
&  =-k_{p}(\rho_{p}\left(  t\right)  )^{-1}S(\rho_{p}\left(  t\right)
)^{-1}r_{p}(\xi_{p})\varepsilon_{p}(\xi_{p})\nonumber\\
&  \text{ \ \ }-(\rho_{p}\left(  t\right)  )^{-1}(\dot{\rho}_{p}\left(
t\right)  \xi_{p}+S(\rho_{v}\left(  t\right)  \xi_{v}-\dot{\overline{p}}%
_{0}(t)))\label{eqn:ksi_p_dot_2}\\
\dot{\xi}_{v} &  =h_{v_{A}}(t,\xi)\nonumber\\
&  =-k_{v}(\rho_{v}\left(  t\right)  )^{-1}M^{-1}\varepsilon_{v}(\xi
_{v})-(\rho_{v}\left(  t\right)  )^{-1}(\dot{\rho}_{v}\left(  t\right)
\xi_{v}\nonumber\\
&  \text{ \ \ }-M^{-1}(f(v_{d}+\rho_{v}\left(  t\right)  \xi_{v}%
)+w(t))+\dot{v}_{d})\label{eqn:ksi_v_dot_2}%
\end{align}
where $M=\mathrm{diag}\left(  \left[  m_{i}\right]  _{i=1,\ldots,N}\right)  $
and $f(v_{d}+\rho_{v}\left(  t\right)  \xi_{v})=\left[  f_{1}(v_{d_{1}}%
+\rho_{v_{1}}\left(  t\right)  \xi_{v_{1}}),\cdots,f_{N}(v_{d_{N}}+\rho
_{v_{N}}\left(  t\right)  \xi_{v_{N}}\right]  ^{T}$ with $m_{i}$, $f_{i}%
(\cdot)$, $i=1,\ldots,N$ denoting the unknown masses and nonlinearities of the
vehicle model (\ref{eqn:nonlinear_dynamics}) respectively. Thus, the closed
loop dynamical system of $\xi(t)=\left[  \xi_{p}^{T}(t),\xi_{v}^{T}(t)\right]
^{T}$ may be written in compact form as:
\begin{equation}
\dot{\xi}=h_{A}(t,\xi)\triangleq%
\begin{bmatrix}
h_{p_{A}}(t,\xi)\\
h_{v_{A}}(t,\xi)
\end{bmatrix}
.\label{eqn:ksi_dot}%
\end{equation}
Let us also define the open set $\Omega_{\xi}=\Omega_{\xi_{p}}\times
\Omega_{\xi_{v}}\subset\Re^{2N}$ with:
\begin{align}
\Omega_{\xi_{p}} &  =(-\underline{M}_{p_{1}},\overline{M}_{p_{1}})\times
\cdots\times(-\underline{M}_{p_{N}},\overline{M}_{p_{N}})\nonumber\\
\Omega_{\xi_{v}} &  =\underbrace{(-1,1)\times\cdots\times(-1,1)}%
_{N\text{-times}}\text{.}\label{eqn:omega}%
\end{align}

\textit{Phase I. }Selecting the parameters $\underline{M}_{p_{i}},$
$\overline{M}_{p_{i}},$ $i=1,\ldots,N$ according to (\ref{eqn:M_upper+lower}),
we guarantee that the set $\Omega_{\xi}$ is nonempty and open. Moreover, owing
to \textbf{Assumption A1}, $\xi_{p}(0)\in\Omega_{\xi_{p}}$, as shown in
(\ref{eqn:pp_t_zero}). Furthermore, selecting $\rho_{v_{i}}\left(  0\right)
>\left\vert e_{v_{i}}(0)\right\vert ,$ $i=1,\ldots,N$ ensures that $\xi
_{v}(0)\in\Omega_{\xi_{v}}$ as well. Thus, we conclude that $\xi(0)\in
\Omega_{\xi}$. Additionally, $h_{A}$ is continuous on $t$ and locally
Lipschitz on $\xi$ over the set $\Omega_{\xi}$. Therefore, the hypotheses of
Theorem 54 in \cite{1998Sontag} (p.p. 476) hold and the existence of a maximal
solution $\xi(t)$ of (\ref{eqn:ksi_dot}) for a time interval $[0,\tau_{\max})$
such that $\xi(t)\in\Omega_{\xi},$ $\forall t\in\lbrack0,\tau_{\max})$ is guaranteed.

\textit{Phase II-Kinematics. }We have proven in \textit{Phase I} that
$\xi(t)\in\Omega_{\xi},$ $\forall t\in\lbrack0,\tau_{\max})$ and more
specifically that:
\begin{equation}
\left.
\begin{array}
[c]{l}%
\xi_{p_{i}}\left(  t\right)  =\frac{e_{p_{i}}(t)}{\rho_{p_{i}}(t)}%
\in(-\underline{M}_{p_{i}},\text{ }\overline{M}_{p_{i}})\\
\xi_{v_{i}}\left(  t\right)  =\frac{e_{v_{i}}(t)}{\rho_{v_{i}}(t)}\in(-1,1)
\end{array}
\right\}  \text{ }i=1,\dots,N \label{eqn:ksi}%
\end{equation}
for all $t\in\lbrack0,\tau_{\max})$, from which we obtain that $e_{p_{i}}(t)$
and $e_{v_{i}}(t)$ are absolutely bounded by $\max\{\underline{M}_{p_{i}},$
$\overline{M}_{p_{i}}\}\rho_{p_{i}}(t)$ and $\rho_{v_{i}}(t)$ respectively for
$i=1,\dots,N.$ Furthermore, owing to (\ref{eqn:ksi}), the error vector
$\varepsilon_{p}(t)$, as defined in (\ref{eqn:epsilon_p}), is well defined for
all $t\in\lbrack0,\tau_{\max})$. Therefore, consider the positive definite and
radially unbounded function $V_{p_{A}}=\frac{1}{2}\varepsilon_{p}%
^{T}P\varepsilon_{p}$, where $P\triangleq($\textrm{diag}$(S^{-1}%
\underline{\mathbf{1}}))^{-1}$ is a diagonal positive definite matrix
satisfying $PS+S^{T}P>0$, as dictated by Lemma \ref{Lemma;M-matrix}.
Differentiating $V_{p_{A}}$ with respect to time, substituting (\ref{eqn:r_p}%
), (\ref{eqn:ksi_p_dot_2}) and exploiting: i) the diagonality of the matrices
$P$, $r_{p}(\xi_{p})$, $\rho_{p}\left(  t\right)  $, ii) the positive
definiteness of $Q\triangleq PS+S^{T}P$ as well as iii) the boundedness of
$\dot{\rho}_{p}\left(  t\right)  $, $\rho_{v}\left(  t\right)  $ and
$\dot{\overline{p}}_{0}(t)$, we get:
\begin{align*}
\dot{V}_{p_{A}}  &  \leq-k_{p}\lambda_{\min}(Q)\left\Vert \varepsilon_{p}%
^{T}r_{p}(\xi_{p})(\rho_{p}\left(  t\right)  )^{-1}\right\Vert ^{2}\\
&  +\left\Vert \varepsilon_{p}^{T}r_{p}(\xi_{p})(\rho_{p}\left(  t\right)
)^{-1}\right\Vert \bar{F}_{p}%
\end{align*}
where $F_{p}$ is a positive constant independent of $\tau_{\max}$,
satisfying:
\begin{equation}
\left\Vert P(\dot{\rho}_{p}\left(  t\right)  \xi_{p}+S(\rho_{v}\left(
t\right)  \xi_{v}-\dot{\overline{p}}_{0}(t)))\right\Vert \leq\bar{F}_{p}
\label{eqn:F_p_inequality}%
\end{equation}
for all $(\xi,t)\in\Omega_{\xi}\times\Re_{+}$. Therefore, we conclude that
$\dot{V}_{p_{A}}$ is negative when $\left\Vert \varepsilon_{p}^{T}r_{p}%
(\xi_{p})(\rho_{p}\left(  t\right)  )^{-1}\right\Vert >\frac{\bar{F}_{p}%
}{k_{p}\lambda_{\min}(Q)}$, from which, owing to the positive definiteness and
diagonality of $r_{p}(\xi_{p})(\rho_{p}\left(  t\right)  )^{-1}$ as well as
employing (\ref{eqn:r_p}) and (\ref{eqn:performance_functions}), it can be
easily verified that:
\begin{equation}
\left\Vert \varepsilon_{p}(t)\right\Vert \leq\bar{\varepsilon}_{p}%
:=\tfrac{\lambda_{\max}(P)}{\lambda_{\min}(P)}\max\left\{  \left\Vert
\varepsilon_{p}(0)\right\Vert ,\tfrac{\bar{F}_{p}\max\left\{  \tfrac
{\underline{M}_{p_{i}}\overline{\text{ }M}_{p_{i}}}{\underline{M}_{p_{i}%
}+\overline{\text{ }M}_{p_{i}}}\right\}  }{k_{p}\lambda_{\min}(Q)}\right\}
\label{eqn:e_p_less_e_p_bar}%
\end{equation}
for all $t\in\lbrack0,\tau_{\max})$. Furthermore, from (\ref{eqn:epsilon_p}),
taking the inverse logarithm, we obtain:%
\begin{multline}
{\small -}\underline{M}_{p_{i}}{\small <-}\tfrac{\exp\left(  \bar{\varepsilon
}_{p}\right)  -1}{\exp\left(  \bar{\varepsilon}_{p}\right)  +\frac
{\underline{M}_{p_{i}}}{\overline{M}_{p_{i}}}}\underline{M}_{p_{i}}%
{\small =}\underline{\xi}_{p_{i}}{\small \leq\xi}_{p_{i}}{\small (t)}\\
{\small \leq\bar{\xi}}_{p_{i}}{\small =}\tfrac{\exp\left(  \bar{\varepsilon
}_{p}\right)  -1}{\exp\left(  \bar{\varepsilon}_{p}\right)  +\frac
{\overline{M}_{p_{i}}}{\underline{M}_{p_{i}}}}\overline{M}_{p_{i}}%
{\small <}\overline{M}_{p_{i}}\label{eqn:ksi_p_final}%
\end{multline}
for all $t\in\lbrack0,\tau_{\max})$ and $i=1,\dots,N$. Thus, the reference
velocity vector $v_{d}(\xi_{p},t),$ as designed in
(\ref{eqn:v_d_predecessor_f}), remains bounded for all $t\in\lbrack
0,\tau_{\max})$. Moreover, invoking $v_{i}\equiv v_{d_{i}}+\rho_{v_{i}}\left(
t\right)  \xi_{v_{i}}$ we also conclude the boundedness of the velocities
$v_{i}(t),$ $i=1,\dots,N$ for all $t\in\lbrack0,\tau_{\max})$. Finally,
differentiating $v_{d}(\xi_{p},t)$ with respect to time, substituting
(\ref{eqn:ksi_p_dot_2}) and utilizing (\ref{eqn:ksi_p_final}), it is
straightforward to deduce the boundedness of $\dot{v}_{d}$ for all
$t\in\lbrack0,\tau_{\max})$ as well.

\textit{Phase II-Dynamics. }Owing to (\ref{eqn:ksi}), the error vector
$\varepsilon_{v}(t)$ (see (\ref{eqn:epsilon_v})) is well defined for all
$t\in\lbrack0,\tau_{\max})$. Therefore, consider the positive definite and
radially unbounded function $V_{v_{A}}=\frac{1}{2}\varepsilon_{v}%
^{T}M\varepsilon_{v}$ where $M=\mathrm{diag}\left(  \left[  m_{i}\right]
_{i=1,\ldots,N}\right)  $ with $m_{i}$, $i=1,\dots,N$ denoting the unknown
mass of the vehicle model (\ref{eqn:nonlinear_dynamics}). Following the same
line of proof with $V_{p_{A}}$ in \textit{Phase II-Kinematics}, we conclude
that:
\begin{equation}
\left\Vert \varepsilon_{v}(t)\right\Vert \leq\bar{\varepsilon}_{v}%
:=\tfrac{\max\{m_{i}\}}{\min\{m_{i}\}}\max\left\{  \left\Vert \varepsilon
_{v}(0)\right\Vert ,\tfrac{\bar{F}_{v}}{2k_{v}}\right\}
\label{eqn:e_v_less_e_v_bar2}%
\end{equation}
for all $t\in\lbrack0,\tau_{\max})$, where $\bar{F}_{v}$ is a positive
constant satisfying:
\begin{equation}
\left\Vert (M\dot{\rho}_{v}\left(  t\right)  \xi_{v}-(f(v_{d}+\rho_{v}\left(
t\right)  \xi_{v})+w(t))+\dot{v}_{d})\right\Vert \leq\bar{F}_{v}
\label{eqn:F_v_inequality}%
\end{equation}
owing to: i) the boundedness of $v_{d}$ and $\dot{v}_{d}$ that was proven
previously, ii) the continuity of function $f_{i}(\cdot)$ and iii) the
boundedness of $\dot{\rho}_{v}\left(  t\right)  $, $\rho_{v}\left(  t\right)
$ as well as of the disturbance term $w(t)$. Furthermore, from
(\ref{eqn:epsilon_v}), taking the inverse logarithmic function, we obtain:%
\begin{equation}
{\small -}1{\small <-}\tfrac{\exp\left(  \bar{\varepsilon}_{v}\right)
-1}{\exp\left(  \bar{\varepsilon}_{v}\right)  +1}{\small =}\underline{\xi
}_{v_{i}}{\small \leq\xi}_{v_{i}}{\small (t)\leq\bar{\xi}}_{v_{i}}%
{\small =}\tfrac{\exp\left(  \bar{\varepsilon}_{v}\right)  -1}{\exp\left(
\bar{\varepsilon}_{v}\right)  +1}{\small <1} \label{eqn:ksi_v_final}%
\end{equation}
for all $t\in\lbrack0,\tau_{\max})$ and $i=1,\dots,N$, which also leads to the
boundedness of the distributed control protocol (\ref{eqn:u}).

\textit{Phase III. }Up to this point, what remains to be shown is that
$\tau_{\max}$ can be extended to $\infty$. In this direction, notice by
(\ref{eqn:ksi_p_final}) and (\ref{eqn:ksi_v_final}) that $\xi(t)\in\Omega
_{\xi}^{^{\prime}}\triangleq\Omega_{\xi_{p}}^{^{\prime}}\times\Omega_{\xi_{v}%
}^{^{\prime}}$, $\forall t\in\lbrack0,\tau_{\max})$, where $\Omega_{\xi_{p}%
}^{^{\prime}}\triangleq\left[  \underline{\xi}_{p_{1}},{\small \bar{\xi}%
}_{p_{1}}\right]  \times\cdots\times\left[  \underline{\xi}_{p_{N}%
},{\small \bar{\xi}}_{p_{N}}\right]  $ and $\Omega_{\xi_{v}}^{^{\prime}%
}\triangleq\left[  \underline{\xi}_{v_{1}},{\small \bar{\xi}}_{v_{1}}\right]
\times\cdots\times\left[  \underline{\xi}_{v_{N}},{\small \bar{\xi}}_{v_{N}%
}\right]  $ are nonempty and compact subsets of $\Omega_{\xi_{p}}$ and
$\Omega_{\xi_{v}}$ respectively. Hence, assuming $\tau_{\max}<\infty$ and
since $\Omega_{\xi}^{^{\prime}}\subset\Omega_{\xi}$, Proposition C.3.6 in
\cite{1998Sontag} (p.p. 481) dictates the existence of a time instant
$t^{^{\prime}}\in\lbrack0,\tau_{\max})$ such that $\xi(t^{^{\prime}}%
)\notin\Omega_{\xi}^{^{\prime}}$, which is a clear contradiction. Therefore,
$\tau_{\max}$ is extended to $\infty$. Thus, all closed loop signals remain
bounded and moreover $\xi(t)\in\Omega_{\xi}^{^{\prime}}\subset\Omega_{\xi}$,
$\forall t\geq0$. Finally, multiplying (\ref{eqn:ksi_p_final}) by $\rho
_{p_{i}}(t)$, $i=1,\dots,N$, we also conclude that:
\begin{equation}
-\underline{M}_{p_{i}}\rho_{p_{i}}\left(  t\right)  <e_{p_{i}}\left(
t\right)  <\overline{M}_{p_{i}}\rho_{p_{i}}\left(  t\right)  \text{, }\forall
t\geq0 \label{eqn:e_p_final}%
\end{equation}
for all $i=1,\dots,N$ and consequently the solution of the robust formation
control problem with prescribed performance under collision and connectivity
constraints for the considered platoon of vehicles. \hfill$\blacksquare$

\begin{remark}
From the aforementioned proof it can be deduced that the proposed control
schemes achieve their goals without resorting to the need of rendering the
ultimate bounds $\bar{\varepsilon}_{p}$, $\bar{\varepsilon}_{v}$ of the
modulated position and velocity errors $\varepsilon_{p}\left(  \xi_{p}\left(
t\right)  \right)  $, $\varepsilon_{v}\left(  \xi_{v}\left(  t\right)
\right)  $ arbitrarily small by adopting extreme values of the control gains
$k_{p}$ and $k_{v}$ (see (\ref{eqn:e_p_less_e_p_bar}) and
(\ref{eqn:e_v_less_e_v_bar2})). More specifically, notice that
(\ref{eqn:ksi_p_final}) and (\ref{eqn:ksi_v_final}) hold no matter how large
the finite bounds $\bar{\varepsilon}_{p}$, $\bar{\varepsilon}_{v}$ are. In the
same spirit, large uncertainties involved in the vehicle nonlinear model
(\ref{eqn:nonlinear_dynamics}) can be compensated, as they affect only the
size of $\bar{\varepsilon}_{v}$ through $\bar{F}_{v}$ (see
(\ref{eqn:F_v_inequality})), but leave unaltered the achieved stability
properties. Hence, the actual performance given in (\ref{eqn:e_p_final}),
which is solely determined by the designer-specified performance functions
$\rho_{p_{i}}\left(  t\right)  $ and the parameters $-\underline{M}_{p_{i}}$,
$\overline{M}_{p_{i}}$, $i=1,\dots,N$, becomes isolated against model
uncertainties, thus extending greatly the robustness of the proposed control schemes.
\end{remark}

\begin{figure*}[t]
\begin{multicols}{2}
\includegraphics[trim =0.1cm 0 0 0,width = 0.50\textwidth, height = 0.23\textheight]{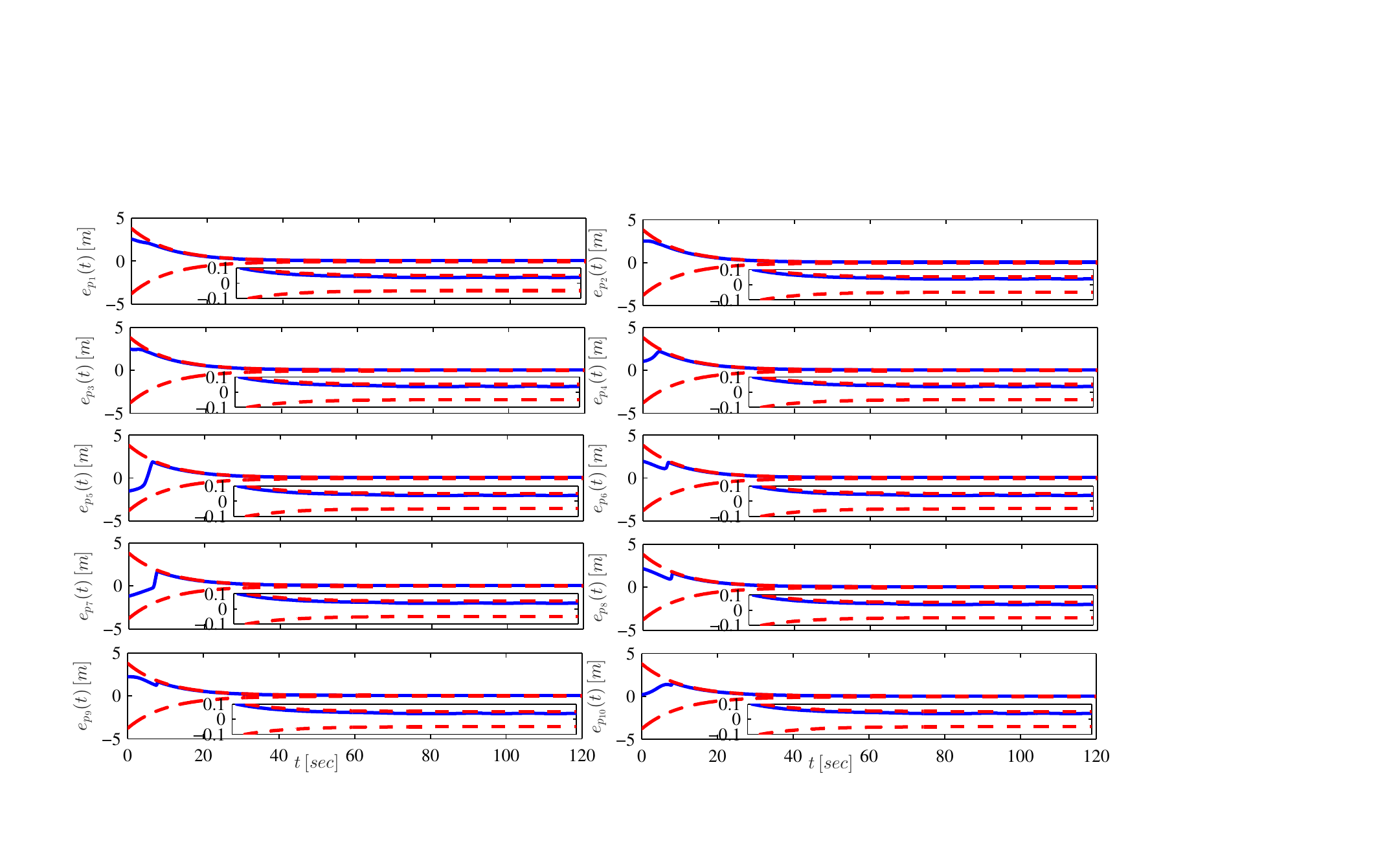}\caption{The position errors $e_{p_{i}}(t)$ (PF). }%
\label{fig:errors_pf}%
\includegraphics[trim =0.44cm 0 0 0,width = 0.49\textwidth, height = 0.23\textheight]{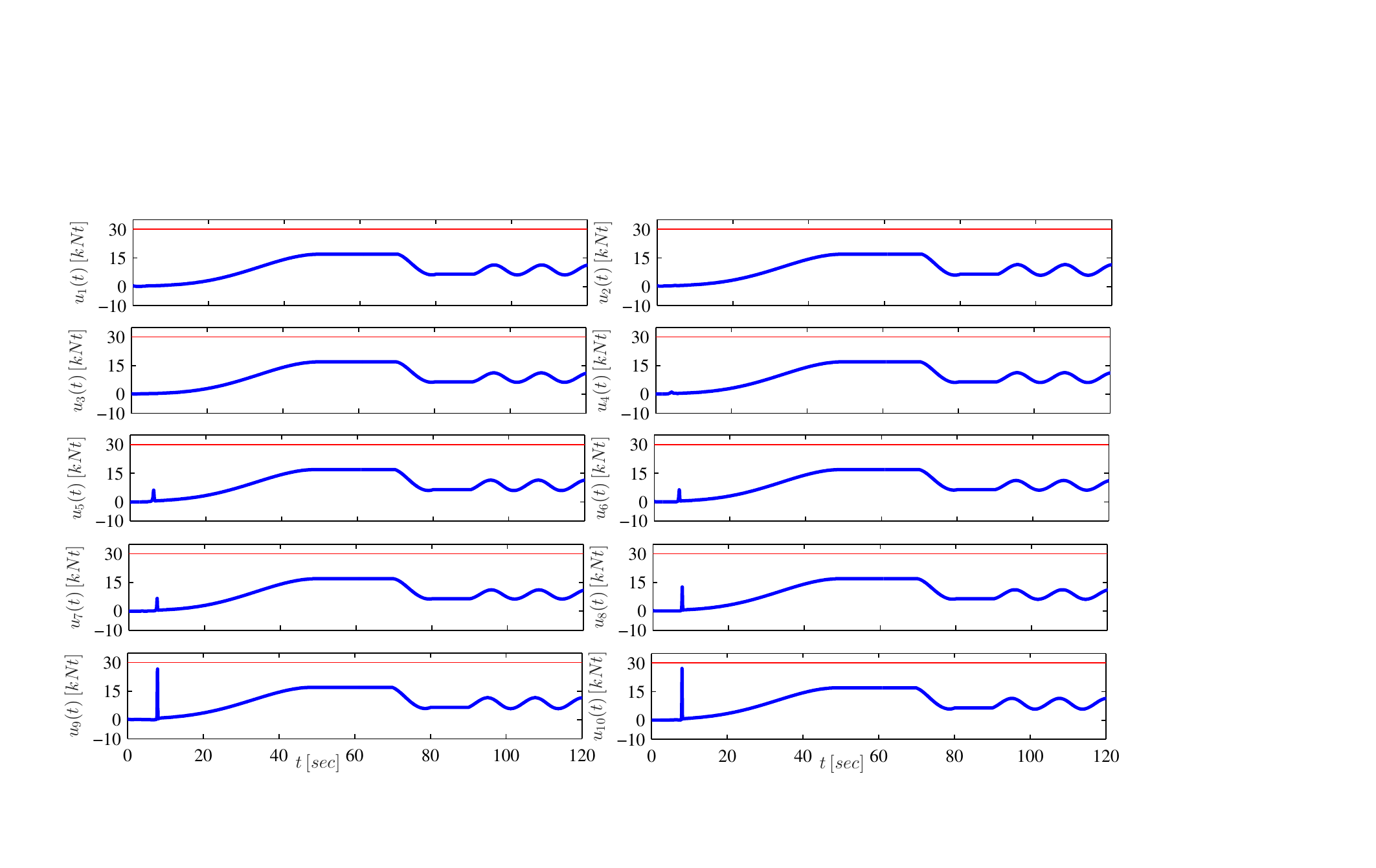}\caption{The required control input signals (PF). }%
\label{fig:inputs_pf}%
\end{multicols}
\vspace{-0.75cm}
\end{figure*}

\begin{remark}
\label{remark_5} (Selecting the Control Gains) It should be noted that the selection of the control gains
affects both the quality of evolution of the neighborhood errors $e_{p_{i}%
}\left(  t\right)  $, $i=1,\dots,N$ inside the corresponding performance
envelopes as well as the control input characteristics (e.g., decreasing the
gain values leads to increased oscillatory behaviour within the prescribed
performance envelope described by (\ref{eqn:e_p_final}), which is improved
when adopting higher values, enlarging, however, the control effort both in
magnitude and rate). Additionally, fine tuning might be needed in real-time
scenarios, to retain the required control input signals within the feasible
range that can be implemented by the actuators. Similarly, the control input
constraints impose an upper bound on the required speed of convergence of
$\rho_{p_{i}}\left(  t\right)  $, $i=1,\dots,N$, as obtained by the
exponentials $\exp\left(  -lt\right)  $. Hence, the selection of the control
gains $k_{p}$ and $k_{v}$ can have positive influence on the overall closed
loop system response. More specifically, notice that (\ref{eqn:F_p_inequality}%
)-(\ref{eqn:ksi_v_final}) provide bounds on $\varepsilon_{p}$ and
$\varepsilon_{v}$ that depend on the constants $\overline{F}_{p}$ and
$\overline{F}_{v}$. Therefore, invoking (\ref{eqn:v_d_predecessor_f}%
),(\ref{eqn:v_d_bidirectional}) and (\ref{eqn:u}) we can select the control
gains $k_{p}$ and $k_{v}$ such that $v_{d}$ and $u$ are retained within
certain bounds. Nevertheless, the constants $\overline{F}_{p}$ and
$\overline{F}_{v}$ involve the parameters of the model, the external
disturbances, the velocity/acceleration of the leader and the desired
performance specifications. Thus, an upper bound of the dynamic parameters of
the system as well as of the exogenous disturbances must be given in order to
extract any relationships between the achieved performance and the input
constraints\footnote{Notice that the proposed methodology does not take explicitly into account any specifications in
the input (magnitude or slew rate). Such research direction is an open issue
for future investigation and would increase significantly the applicability
of the proposed scheme.}. Finally, in the same direction, the selection of the velocity
performance functions $\rho_{v_{i}}\left(  t\right)  $, $i,\dots,N$ affects
both the evolution of the position errors within the corresponding performance
envelopes as well as the control input characteristics.
\end{remark}


\begin{remark}
(String Stability) Note that the proposed algorithm guarantees string stability
for the equilibrium point $e_{p_i} = 0, i=1,\dots,N$, in the sense of \cite{1996Swaroop_Hedrick} (see Def. 1). 
In particular, for any $\epsilon > 0$, we can choose $\delta = \max_i\{\max\{\underline{M}_{p_i},\overline{M}_{p_i}\}\} = \epsilon$, $i=1,\dots,N$.
Then, from the aforementioned analysis it can be deduced that $\max_i\lvert e_{p_i}(0) \rvert < \delta$ implies $\max_i\{  \sup_{t\geq0} \lvert e_{p_i}(t) \rvert  \} < \epsilon$, $i=1,\dots,N$. 
\end{remark}

\section{Simulation Results\label{sec: simulations}}

\subsection{Generic Evaluation\label{sub:General-Evaluation}}

To demonstrate the efficiency of the proposed distributed control protocols,
we considered a platoon of $N=10$ vehicles obeying
(\ref{eqn:nonlinear_dynamics}) with $f_{i}\left(  v_{i}\right)  =-50v_{i}%
-25\left\vert v_{i}\right\vert v_{i}$, $w_{i}\left(  t\right)  =A_{i}%
sin(\omega_{i}t+\phi_{i})$ and $m_{i}$, $A_{i}$, $\omega_{i}$, $\phi_{i}$
randomly selected within $\left[  500,1500\right]  $ kg, $[1.0,\,1.5]$ kNt,
$[2\pi,\,4\pi]$ rad/s and $[0,\,2\pi]$ rad respectively for $i=1,\ldots,10$.
Although the size of the aforementioned intervals affects directly the
magnitude of the control effort $u$, which however can be regulated by tuning
appropriately the gains $k_{p}$ and $k_{v}$, as mentioned in Remark
\ref{remark_5}, in view of the theoretical analysis, the uncertainty of the
aforementioned parameters does not affect the performance of the proposed
schemes. Furthermore, the leading vehicle adopts the following continuous
velocity profile:
\[
v_{0}\left(  t\right)  =\left\{
\begin{array}
[c]{lr}%
\frac{75t^{2}-t^{3}}{2500}, & t\in\left[  0,50\right] \\
25, & t\in\left[  50,70\right] \\
0.02t^{3}-4.5t^{2}+336t-8305, & t\in\left[  70,80\right] \\
15, & t\in\left[  80,90\right] \\
17.5-2.5\cos\left(  \frac{t-90}{2}\right)  , & t\in\left[  90,120\right]
\end{array}
\right.  \
\]
whereas the desired distance between consecutive vehicles is equally set at
$\Delta_{i-1,i}=\Delta^{\star}=4$ m, $i=1,\ldots,10$ with the collision and
connectivity constraints given by $\Delta_{\mathsf{col}}=0.05\Delta^{\star}$
and $\Delta_{\mathsf{con}}=1.95\Delta^{\star}$ respectively. Notice that the
aforementioned formation problem under the collision/connectivity constraints
is feasible since $\Delta_{\mathsf{col}}<\Delta_{i-1,i}<\Delta_{\mathsf{con}}%
$, $i=1,\ldots,10$. Moreover, we require steady state errors of no more than
$0.05$ m and minimum speed of convergence as obtained by the exponential
$\exp\left(  -0.1t\right)  $. Thus, according to
(\ref{eqn:performance_functions}) and (\ref{eqn:M_upper+lower}), we selected
the parameters $\underline{M}_{p_{i}}=\overline{M}_{p_{i}}=0.95\Delta^{\star}$
and the functions $\rho_{p_{i}}(t)=(1-\frac{0.05}{0.95\Delta^{\star}}%
)\exp\left(  -0.1t\right)  +\frac{0.05}{0.95\Delta^{\star}}$, $i=1,\ldots,10$
in order to achieve the desired transient and steady state performance
specifications as well as to comply with the collision and connectivity
constraints. Moreover, we chose $\rho_{v_{i}}(t)=2\left\vert e_{v_{i}%
}(0)\right\vert \exp\left(  -0.1t\right)  +0.1$ in order to satisfy
$\rho_{v_{i}}\left(  0\right)  >\left\vert e_{v_{i}}\left(  0\right)
\right\vert $, $i=1,\ldots,10$. Finally, in view of the desired motion profile
of the leader as well as the masses of the vehicles, we chose the control
gains as $k_{p}=0.1,$ $k_{v}=100$ for the predecessor-following architecture
and $k_{p}=10,$ $k_{v}=1000$ for the bidirectional architecture, to obtain
control inputs that satisfy $\left\vert u_{i}\right\vert \leq30$ kNt,
$i=1,\dots,10$.

\begin{figure*}[t]
\begin{multicols}{2}
\includegraphics[trim =0.1cm 0 0 0,width = 0.5\textwidth, height = 0.23\textheight]{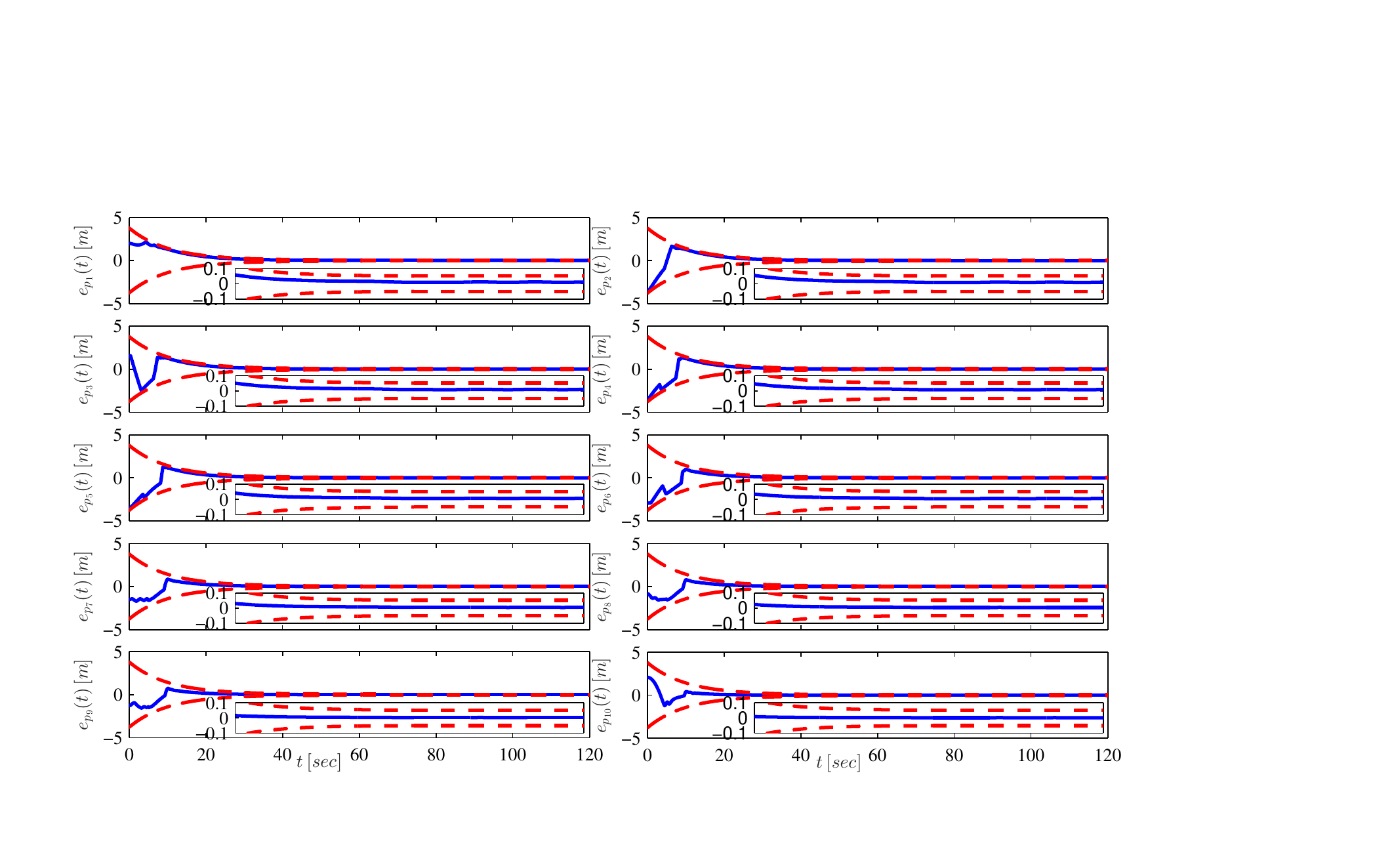}\caption{The position errors $e_{p_{i}}(t)$ (BD). }%
\label{fig:errors_bid}%
\includegraphics[trim =0.42cm 0 0 0,width = 0.49\textwidth, height = 0.23\textheight]{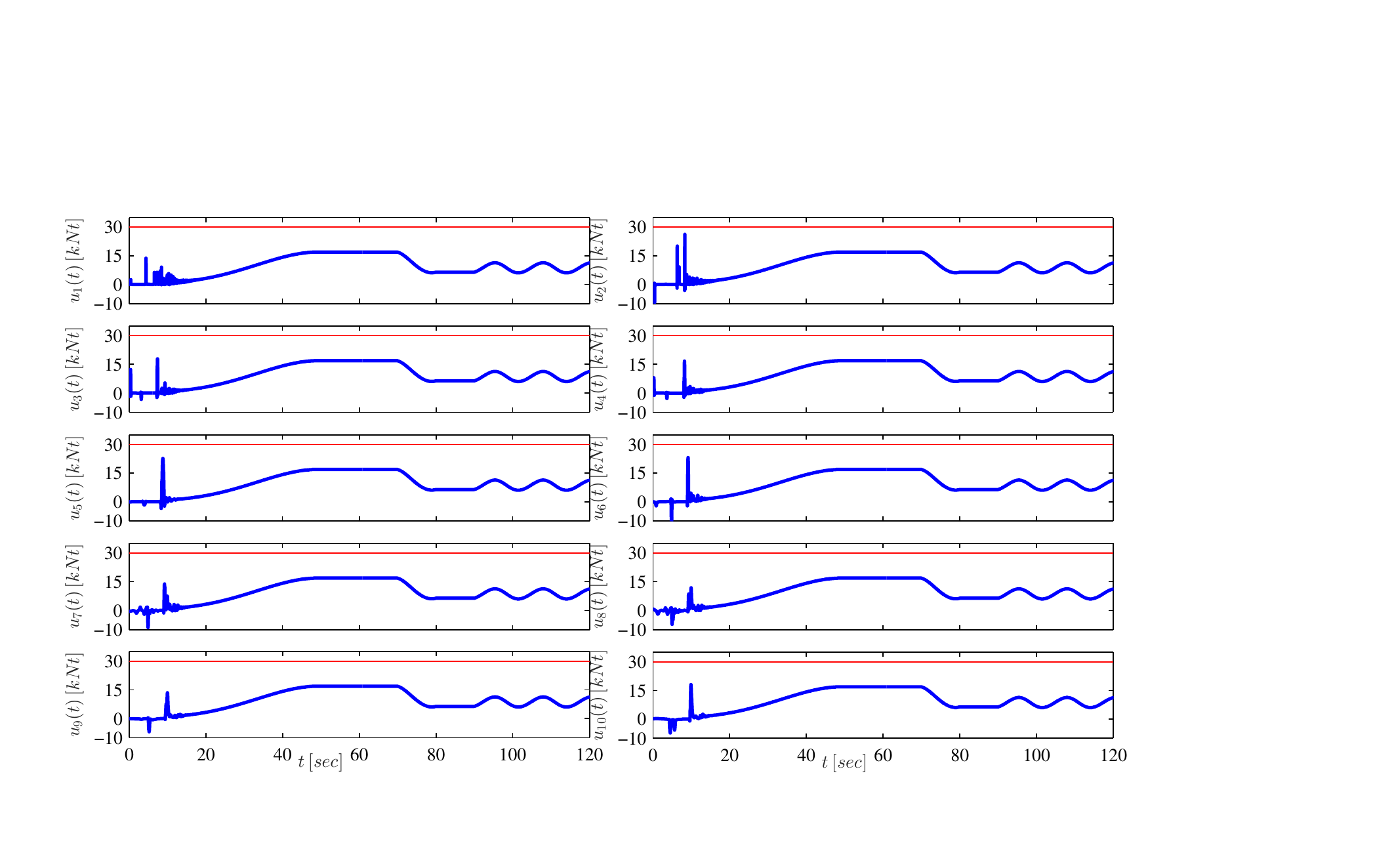}\caption{The required control input signals (BD). }%
\label{fig:inputs_bid}%
\end{multicols}
\vspace{-0.75cm}
\end{figure*}

The simulation results are illustrated in Figs. \ref{fig:errors_pf}%
,\ref{fig:inputs_pf} and \ref{fig:errors_bid},\ref{fig:inputs_bid} for the
predecessor-following (PF) and the bidirectional (BD) control architectures
respectively. More specifically, the evolution of the neighborhood position
errors $e_{p_{i}}(t)$, $i=1,\ldots,10$ along with the corresponding
performance functions are depicted in Figs. \ref{fig:errors_pf} and
\ref{fig:errors_bid}, 
while the required control inputs are illustrated in Figs. \ref{fig:inputs_pf} and \ref{fig:inputs_bid}.
As it was predicted by the theoretical analysis, the formation control problem
with prescribed transient and steady state performance is solved with bounded
closed loop signals for both control architectures, despite the presence of
external disturbances as well as the lack of knowledge of the vehicle dynamic model.



\begin{figure}[h]
\centering\includegraphics[width=\columnwidth]{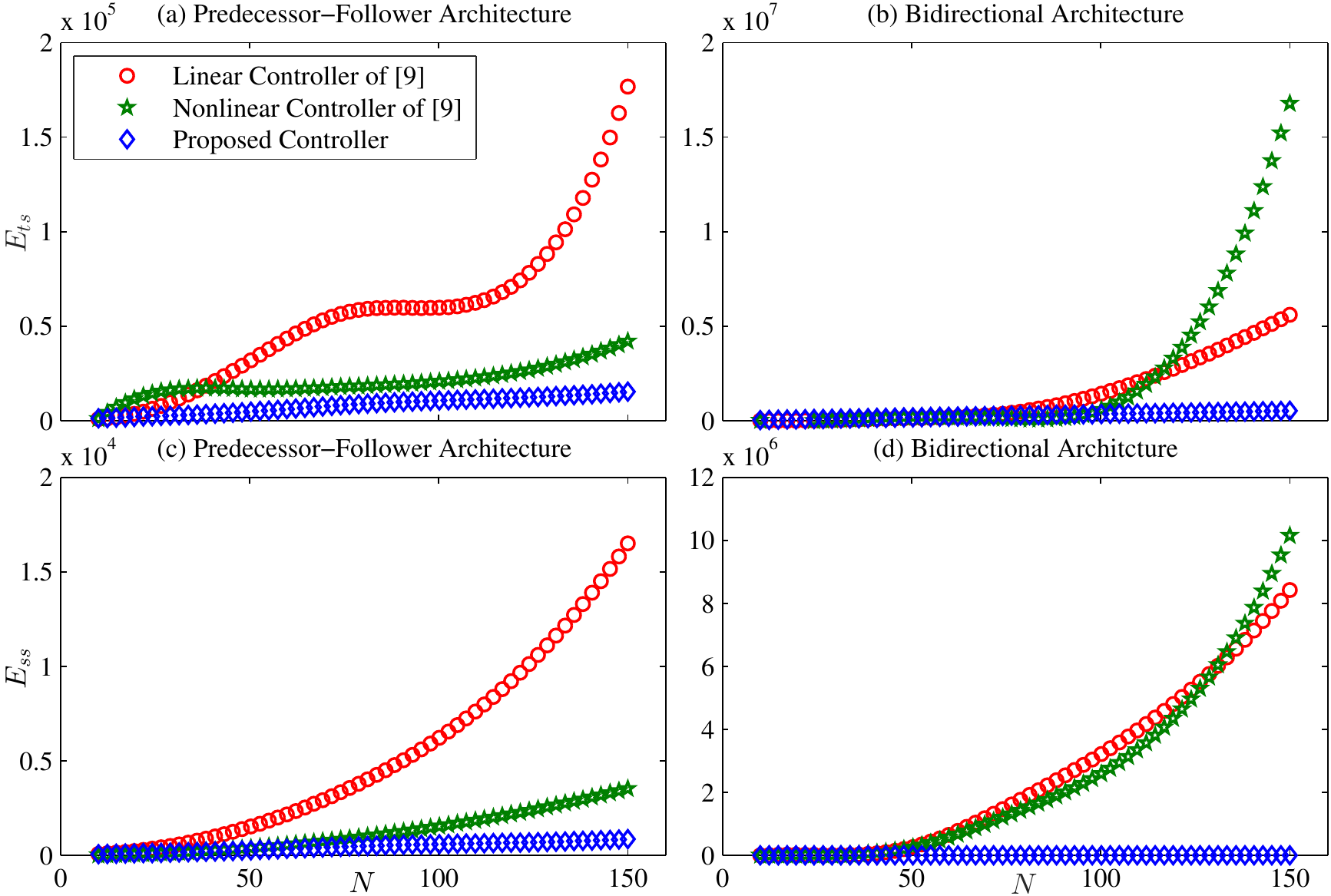} \vspace
{-0.25cm}\caption{The error metrics $E_{ts}$ and $E_{ss}$ for the PF and BD architectures, as the number of vehicles $N$ increases.}%
\label{fig:Comparisons_all}%
\end{figure}

\subsection{Comparative Studies\label{sub:Comparative-Evaluation}}

To investigate further the performance of the proposed methodology, a
comparative simulation study was carried out, on the basis of the
aforementioned nonlinear model, among the proposed control schemes and
the\ linear as well as nonlinear control protocols presented in
\cite{2013Hao_Barooah}. For comparison purposes, we adopted the following
metrics of performance:
\begin{align}
E_{ts}  &  =\frac{1}{N}\int_{0}^{t_{s}}\sum_{i=1}^{N}\{(e_{p_{0,i}}%
(t))^{2}+(\dot{e}_{p_{0,i}}(t))^{2}\}dt\label{eq:metric}\\
E_{ss}  &  =\frac{1}{N}\int_{t_{s}}^{T}\sum_{i=1}^{N}\{(e_{p_{0,i}}%
(t))^{2}+(\dot{e}_{p_{0,i}}(t))^{2}\}dt \label{eq:metric2}%
\end{align}
for the transient and the steady state respectively, where $e_{p_{0,i}}(t)$,
$i=1,\ldots,N$ denote the distance errors with respect to the leader, $t_{s}$
denotes the transient period and $T$ is the overall simulation time. In
particular, we study through extensive numerical simulations how the metrics
$E_{ts}$ and $E_{ss}$ scale with the number of agents $N\in\left[
10,150\right]  $ for $T=120$. It should be noticed that the methods proposed
in \cite{2013Hao_Barooah} considered a double integrator model and therefore a
feedback linearization technique was adopted in the control scheme initially.
However, to simulate a realistic scenario, the model parameters adopted in the
feedback linearization technique deviated up to $15\%$ from their actual
values. Additionally, the corresponding control gains were selected through a
tedious trial-and-error process to yield satisfactory performance for $N=10$.
Regarding the proposed control schemes, the parameters were chosen as in
Section \ref{sub:General-Evaluation}, except for the steady state error bound
and the minimum convergence speed of the performance functions $\rho_{p_{i}%
}\left(  t\right)  ,\rho_{v_{i}}\left(  t\right)  $. In particular, 
$\rho_{\infty}$ was calculated as $\rho_{\infty}=\frac
{0.5\sigma_{\min}\left(  S\right)  }{\sqrt{N}}$, 
and the minimum speed of convergence was
obtained by the exponential $\exp\left(  -2t\right)  $. Finally, the desired
velocity profile of the leader and the desired inter-vehicular
distances were set as in Section \ref{sub:General-Evaluation}.

\begin{figure*}[t]
\begin{multicols}{3}
\includegraphics[width=5.85cm,height=4.0cm]{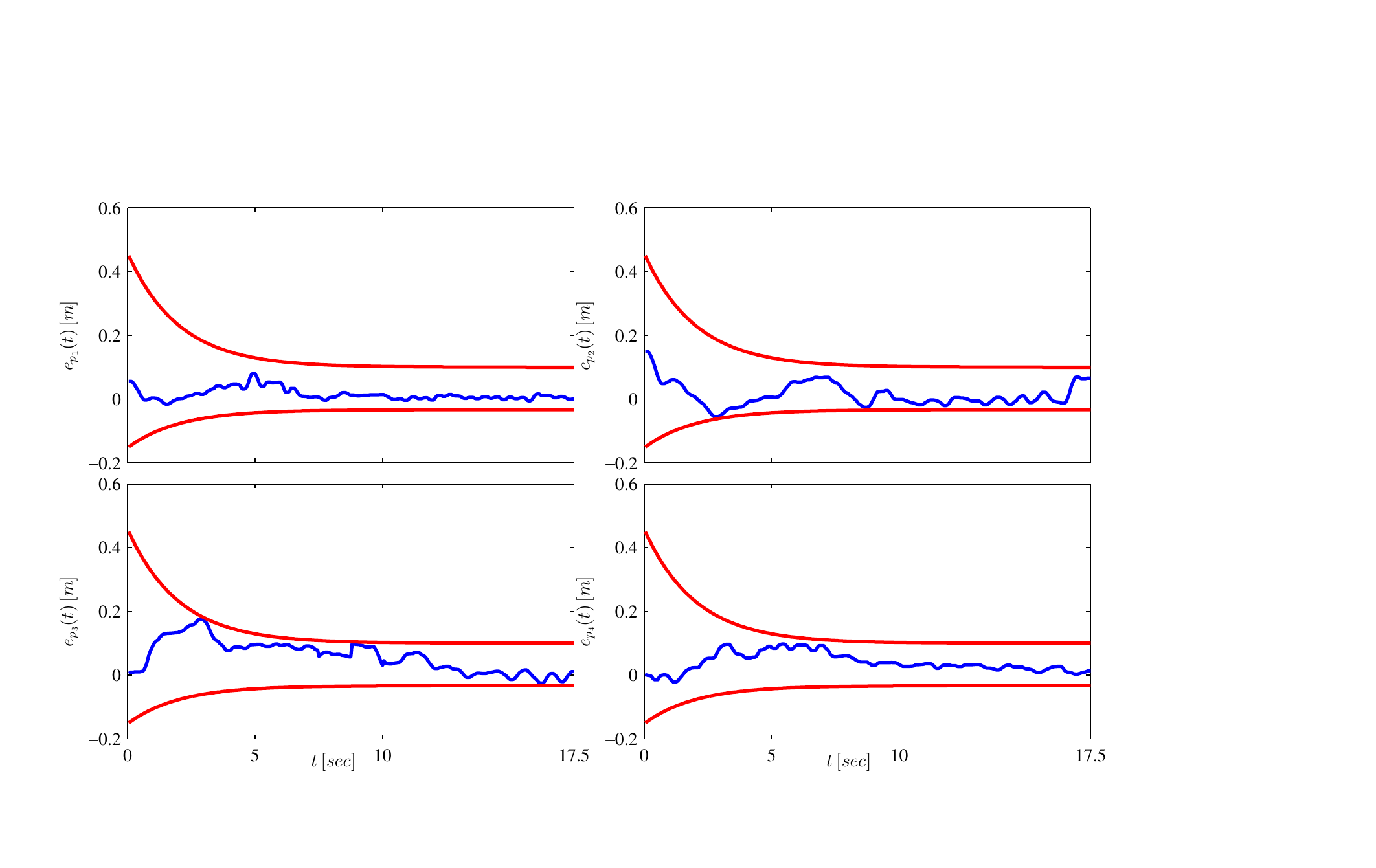}\caption{The position errors $e_{p_{i}}(t),i=1,\dots,4$.}\label{fig:errors_exp}
\includegraphics[width=5.85cm,height=4.0cm]{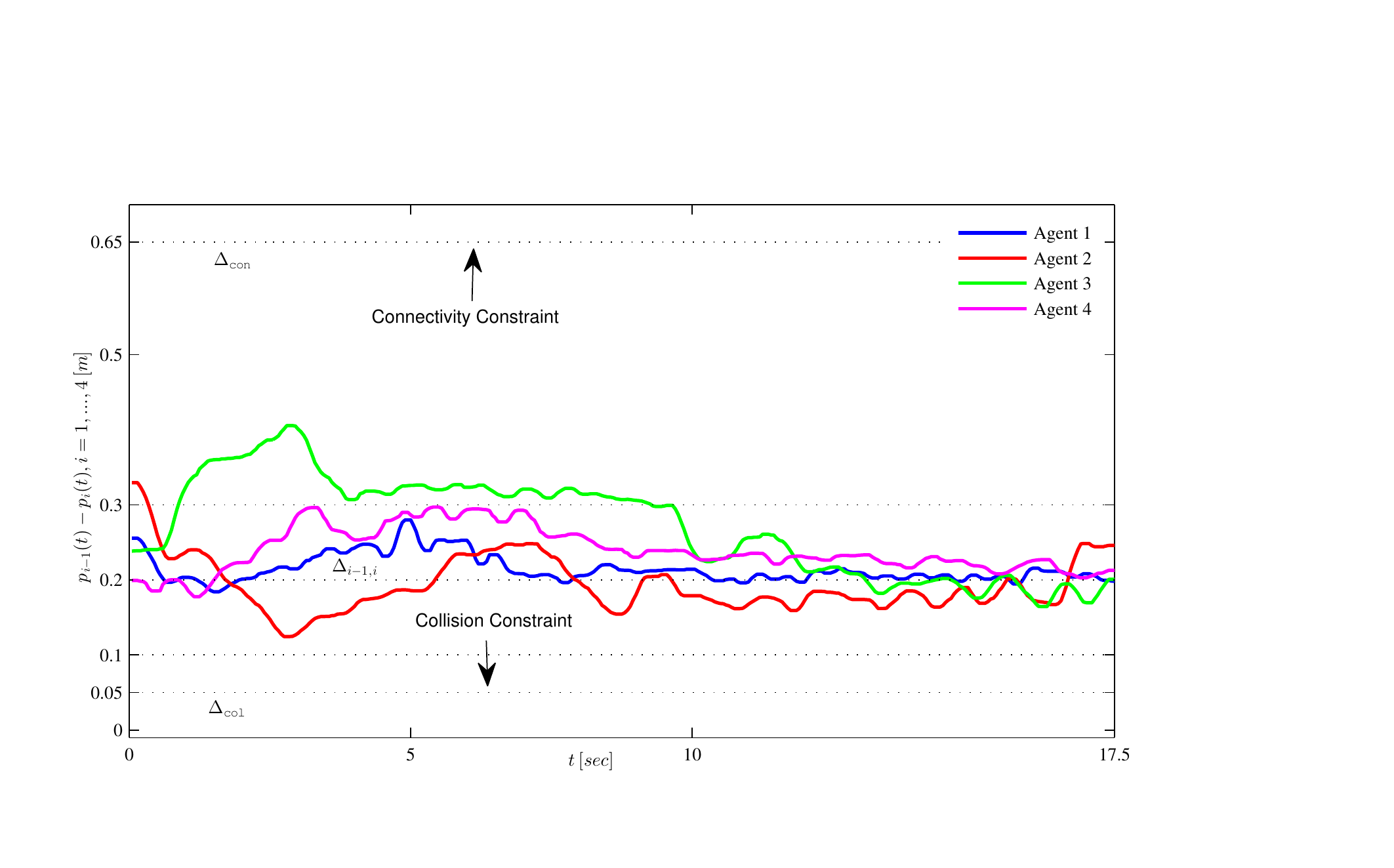}\caption{The distance between successive vehicles.}\label{fig:distances_exp}%
\includegraphics[width=5.85cm,height=4.0cm]{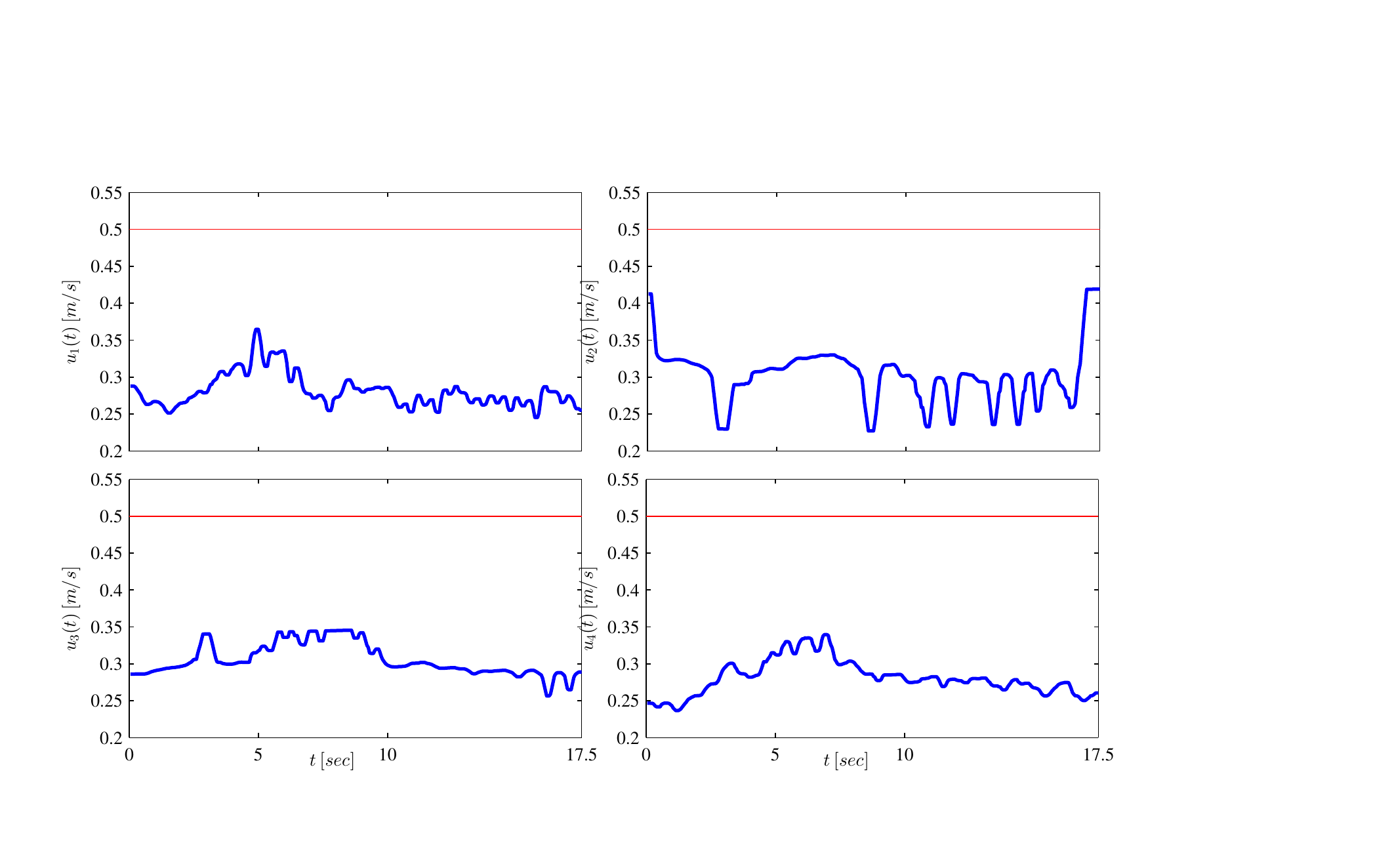}\caption{The required control input signals.}\label{fig:inputs_exp}%
\end{multicols}\vspace{-0.75cm}\end{figure*}

The results of the comparative simulation study are given in Figs.
\ref{fig:Comparisons_all}a-\ref{fig:Comparisons_all}d. More specifically,
Figs. \ref{fig:Comparisons_all}a and \ref{fig:Comparisons_all}b illustrate the
evolution of $E_{ts}$ for the predecessor-following and the bidirectional
control architecture respectively. Similarly, the evolution of $E_{ss}$ is
given in Figs. \ref{fig:Comparisons_all}c and \ref{fig:Comparisons_all}d.
Notice that the proposed control protocols render the metrics $E_{ts}$ and
$E_{ss}$ almost invariant to the number of vehicles $N$. On the contrary, the
performance of the linear and nonlinear control methodologies proposed in
\cite{2013Hao_Barooah} deteriorated in both control architectures as the
number of vehicles increased, proving thus the superiority of the proposed
control protocols.

\section{Experimental Results\label{sec: experiment}}

To verify the performance of the proposed scheme, an experimental procedure
was carried out for the case of the predecessor-following architecture. The
experiment took place along a $10$ m long hallway and lasted approximately
$18$ seconds. Five mobile robots were employed. Particularly, a Pioneer2AT was
assigned as the leading vehicle whereas two KUKA youBot platforms and two
Pioneer2DX mobile robots consisted the following vehicles. To acquire the
inter-vehicular distance measurements, infrared proximity sensors operating
from $5$ to $65$ cm were utilized. The control scheme was designed at the
kinematic level, i.e. the control inputs were the desired velocities
(\ref{eqn:v_d_predecessor_f}) since the embedded motor controller of the
vehicles was responsible for implementing the actual wheel torque commands
that achieved the desired velocities.

The leader adopted a constant velocity model given by $p_{0}(t)=0.3t$ m and
$v_{0}\left(  t\right)  =0.3$ m/s. The desired inter-vehicular distances were
set at $\Delta_{i-1,i}=\Delta^{\star}=0.2$ m, $i=1,\ldots,4$, whereas the
collision and connectivity constraints were given by $\Delta_{\mathsf{col}%
}=0.05$ m and $\Delta_{\mathsf{con}}=0.65$ m respectively, incorporating the
limitations of the infrared sensors. Moreover, we required steady state errors
of no more than $0.1$ m and minimum speed of convergence as obtained by the
exponential $\exp\left(  -0.5t\right)  $. Therefore, we selected
$\underline{M}_{p_{i}}=0.15$ m, $\overline{M}_{p_{i}}=0.45$ m and $\rho
_{p_{i}}(t)=0.78\exp\left(  -0.5t\right)  +0.22$, $i=1,\ldots,4$ in order to
achieve the desired transient and steady state performance specifications as
well as to comply with the collision and connectivity constraints. Finally,
given the maximum velocities of the experimental platforms, we chose
$k_{p}=0.001$ to retain the commanded linear velocities within the range of
velocities $\left\vert v_{i}\right\vert \leq0.5$m/s, $i=1,\dots,4$, that can
be implemented by the embedded motor controllers.

The experimental results are given in Figs. \ref{fig:errors_exp}%
-\ref{fig:inputs_exp}. More specifically, the evolution of the neighborhood
position errors $e_{p_{i}}(t)$, $i=1,\ldots,4$ along with the corresponding
performance functions are depicted in Fig. \ref{fig:errors_exp}. The distance
between subsequent vehicles along with the collision and connectivity
constraints are pictured in Fig. \ref{fig:distances_exp}. The required
velocity commands are illustrated in Fig. \ref{fig:inputs_exp}. It should be
noted that the aforementioned real-time experiment verified the transient and
steady state performance attributes of the proposed distributed control
protocols, despite the sensor inaccuracies and motor limitations, which
constitute the main and most challenging issues compared to computer simulations.

\bibliographystyle{IEEEtran}
\bibliography{JOURNAL_1D_2_ARCHS}

\end{document}